\documentclass{article}
\usepackage[left=3cm,right=3cm,top=3cm,bottom=3cm]{geometry} 
\usepackage{amsmath} 
\usepackage{authblk}
\providecommand{\keywords}[1]{\textbf{\textit{Keywords:}} #1}

\usepackage{graphicx}
\usepackage{amssymb}
\usepackage{amsmath}
\usepackage{multirow}
\usepackage{float}
\usepackage{mathtools}
\mathtoolsset{showonlyrefs=true}
\usepackage{natbib}
\usepackage{rotating}
\usepackage{subfig}
\newcommand{\btheta}{ \mbox{\boldmath $ \theta $} }

\newcommand{\bmu}{ \mbox{\boldmath $\mu$} }

\newcommand{\bSigma}{ \mbox{\boldmath $\Sigma$} }

\newcommand{\atan}{\mbox{atan}}

\usepackage{xcolor}

\begin{document}

\title{Spatio-temporal circular models with non-separable covariance structure
}


\author[1]{Gianluca Mastrantonio}
\author[2]{Giovanna Jona Lasinio}
\author[3]{Alan E. Gelfand}

\affil[1]{ Roma Tre University, Via Silvio D'Amico 77,  Rome, 00145, Italy}
\affil[2]{Sapienza University of Rome, P.le Aldo Moro 5,  Rome, 00185, Italy}
\affil[3]{Duke University,  223-A Old Chemistry Building Box 90251,  Durham, NC 27708-0251, USA}

\date{}

\maketitle

\begin{abstract}
	Circular data arise in many areas of application.  Recently, there has been interest
	in looking at circular data collected separately over time and over space.  Here, we extend some of this work
to the spatio-temporal setting, introducing space-time dependence. We accommodate covariates, implement full kriging and forecasting, and also allow for a nugget which can be time dependent. We work within a Bayesian framework, introducing suitable latent variables to facilitate Markov chain Monte Carlo (MCMC) model fitting. The Bayesian framework enables us to implement full inference, obtaining predictive distributions for kriging and forecasting. We offer comparison between the less flexible but more interpretable wrapped Gaussian process and the more
flexible but less interpretable projected Gaussian process.  We do this illustratively using both simulated data and data from computer model output for wave directions in the Adriatic Sea off the coast of Italy.

\keywords{ Average prediction error \and continuous ranked probability score  \and Kriging  \and Markov chain Monte Carlo \and projected distribution \and wrapped distribution}
\end{abstract}

\section{Introduction}
\label{sec:1}
Circular data, i.e., observations with support on the unit circle, arise in many contexts.
Examples include natural directions, such as wind directions (meteorology), animal movement directions (biology) and rock fracture orientations (geology).
Another type of circular data arises by wrapping periodic time data with period $L$ (say,
day or week) onto a circle with circumference $L$ and then rescaling the circumference to $2\pi$, that of the unit circle. Two dimensional directional data may be observed in space and time, along with linear variables,
as in marine studies where for example wave heights and directions are jointly observed, or in atmospheric modeling where wind fields are represented by wind intensity as well as direction. 
Due to the restriction of the domain to the circle, analysis of circular data must be treated differently
from linear data.  Customary statistical summaries are replaced with their circular counterparts.
For a discussion of inference with circular variables see, e.g., \cite{fisher1996}, \cite{Merdia1999}, \cite{Jammalamadaka2001}
or the recent paper by \cite{lee2010}.

The contribution of this paper is to extend recent spatial and spatio-temporal  circular data models.  In particular,
\cite{Jona2013} consider the use of the wrapped normal approach by developing the wrapped Gaussian process while \cite{wang2014}
consider the use of the projected normal approach by developing the projected Gaussian process.  Here we: (i) extend both processes to the
spatio-temporal setting, introducing space-time dependence, (ii)  introduce space and time varying covariate information,
(iii) show how to implement fully model-based kriging and forecasting, (iv)  allow a nugget which can be time dependent,
and (v)  provide an extensive comparison between the more sparsely parametrized wrapped Gaussian process with the more
flexible projected Gaussian process. We do this illustratively using simulation, as a proof of concept, as well as with data in the form of computer
model output for wave directions in the Adriatic Sea off the coast of Italy.  The models are fitted under a Bayesian framework,
introducing suitable latent variables, enabling full inference.

Modeling of circular data has progressed well beyond the i.i.d. case.   Examples include  linear models \citep[][]{harrison,
fisher1996,Katoa2008}, linear models in a Bayesian context \citep[][]{guttorp1998, damien-walker}, models for circular time
series \citep[][]{breckling1989, Fisher1992,coles98,   Holzmann2006,ravindran}, and hidden Markov models to
address classification issues \citep{lagona2011, Bulla2012, mastrantonio2015}.
In \cite{kato2010}  a Markov process for circular variables is presented.  \cite{Jona2013} consider a spatial wrapped
Gaussian process. \cite{Wang2013}   explore the general projected normal model
while in \cite{wang2014} 
Bayesian analysis of space-time circular data is developed using  projected Gaussian processes. In \cite{wang:2014}
directional wave data is modeled jointly with linear wave height data.

The format of the remainder of the paper is as follows.  In Section \ref{sec:2} we review the wrapping approach and offer a non-separable space-time model for circular data.  In Section \ref{sec:projM}, an analogous model is presented  using the projected normal process. Section \ref{sec:sim} presents several simulation examples giving insight into the inferential performance of the models, while Section \ref{sec:real} analyzes the behavior of the models for wave directions. Section \ref{sec:ext} extends the modeling approach to enable space-time varying covariates reflecting sea state at a location and time.
Some concluding remarks are provided in Section \ref{sec:conc}. Implementation details, further simulated examples and more details on the real data application are
available in the Supplementary Online Material, Sections S1, S2 and S3.

\section{A brief review of the wrapped modeling approach}
\label{sec:2}

Let $Y \in \mathbb{R}$ be a random variable on the real line and let $g(y)$ and $G(y)$ be respectively its  probability density function and cumulative distribution function. The random  variable
\begin{equation}
X=Y\hbox{\rm mod }2\pi, \,  0 \leq X < 2\pi  \label{wrap}
\end{equation}%
is the wrapped version of $Y$ having period $2 \pi$. The probability density function of $X$, $f(x)$, is obtained by wrapping the probability density function of $Y$, $g(y)$, around a circle of unit radius via the transformation $Y$=$ X+2\pi K $, with $K
\in \mathbb{Z}\equiv \{0, \pm 1, \pm 2,\ldots\}$, and takes the form
\begin{equation}
f(x)=\sum_{k=-\infty }^{\infty }{g(x+2\pi k  )}, \label{eq:Eq.1.45}
\end{equation} %
that is, a doubly infinite  sum.

Equation \eqref{eq:Eq.1.45} shows that $g(x+2\pi k)$ is the joint distribution of $(X,K)$.  Hence, the marginal distribution of
$K$ is $P(K=k) = \int_{0}^{2\pi} g(x+2\pi k) dx$,  the  conditional distributions $P(K=k|X=x) = g(x+2\pi k)/\sum_{j=
-\infty}^{\infty} g(x+2\pi j)$ and the  distribution of $X|K=k$ is $g(x+2\pi k)/\int_{0}^{2\pi} g(x+2\pi  k) dx$.
The introduction of  $K$ as latent variable facilitates model fitting \citep{Jona2013}.

Following \cite{coles98}, we can extend the wrapping approach to multivariate distributions.  Let $\mathbf{Y} =
(Y_{1},Y_{2},\ldots,Y_{p})\sim g(\cdot)$, with  $g(\cdot)$  a $p-$variate distribution on $\mathbb{R}^{p}$ indexed by say
$\boldsymbol{\theta}$ and let $\mathbf{K} = (K_{1},K_{2},\ldots,K_{p})$ be such that $\mathbf{Y}= \mathbf{X} + 2\pi \mathbf{K}$.
Then the distribution of $\mathbf{X}$ is
\begin{equation}\label{GMVW}
	f(\mathbf{x}) = \sum_{k_{1}= -\infty}^{+\infty}\sum_{k_{2}= -\infty}^{+\infty}\ldots\sum_{k_{p}= -\infty}^{+\infty} g(\mathbf{x}+ 2 \pi \mathbf{k}).
\end{equation}

From \eqref{GMVW} we see, as in the univariate case, that the joint density of $(\mathbf{X}, \mathbf{K})$ is $g(\mathbf{x}+ 2 \pi \mathbf{k})$.
If $g(\cdot; \btheta)$ is a $p$-variate normal density, with $\btheta = (\bmu, \bSigma)$, then $\mathbf{X}$ has a $p$-variate
wrapped normal  distribution  with  parameters $(\bmu, \bSigma)$. Here, we introduce the latent random vector of winding
numbers $\mathbf{K}$ to facilitate model fitting. \cite{Merdia1999}  point out that only a few values of $K$ are needed to obtain a reasonable approximation of the wrapped distribution  and \cite{Jona2013}  show, when $g(\cdot; \btheta)$ is Gaussian, how to choose the set of values of  $K$ based on the variance of the associated conditional distribution.

Let $Y(\mathbf{s})$ be a Gaussian process (GP) with $\mathbf{s} \in \mathbb{R}^2$, mean function $\mu(\mathbf{s})$ and covariance function
say $\sigma^{2} \rho(||\mathbf{s}_i-\mathbf{}s _j ||; \boldsymbol{\psi})$, where $\boldsymbol{\psi}$ is a set of parameters. For a set of locations $\mathbf{s}_{1},\mathbf{s}_{2},\ldots,\mathbf{s}_{n}$,
$\mathbf{Y}=(Y(\mathbf{s}_{1}),Y(\mathbf{s}_{2}),\ldots,Y(\mathbf{s}_{n})) \sim N(\boldsymbol{\mu}, \sigma^2
C(\boldsymbol{\psi}))$, where   $\boldsymbol{\mu}= (\mu(\mathbf{s}_{1}),\ldots, \mu(\mathbf{s}_{n}))$ and
$C(\boldsymbol{\psi})_{ij}= \rho(\mathbf{s}_{i} - \mathbf{s}_{j};\boldsymbol{\psi})$.  As a consequence
$\mathbf{X}=(X(\mathbf{s}_{1}),X(\mathbf{s}_{2}),\ldots,X(\mathbf{s}_{n})) \sim
WrapN(\boldsymbol{\mu},\sigma^{2}\mathbf{C}(\boldsymbol{\psi}))    $ \citep{Jona2013}, where  $WrapN(\cdot,\cdot)$ indicates the wrapped normal distribution.

\subsection{Space-time model specification and model fitting}\label{sec:WNmodfit}

Turning to space and time, suppose we seek  $\{ X(\mathbf{s},t) \in [0,2 \pi), \mathbf{s} \in \mathcal{S} \subseteq \mathbb{R}^2
, t\in \mathcal{T} \subseteq \mathbb{Z}^+  \}$, a  spatio-temporal process of angular variables.
We can  model  $X(\mathbf{s},t) $  as a  spatio-temporal wrapped Gaussian process  through its linear counterpart
$Y(\mathbf{s},t)$, extending the above approach.
We assume that the linear process is a spatio-temporal Gaussian process  having   non-separable covariance
structure with variance $\sigma^2$ and the  stationary correlation function due to Gneiting  (see equation (14) in \cite{Gneiting2002}):
\begin{equation}\label{eq:cov1}
	\mbox{Cor}(Y(\mathbf{s},t), Y(\mathbf{s}',t'))\equiv \rho(\mathbf{h},u)=\frac{1}{(a|u|^{2\alpha}+1)^{ \tau}}\exp\left(-\frac{c\|\mathbf{h}\|^{2\gamma}}{(a|u|^{2\alpha}+1)^{\beta\gamma}}\right),
\end{equation}
where $(\mathbf{h},u)\in\mathbb{R}^d\times\mathbb{R}$,  $\mathbf{h} =  \mathbf{s}-\mathbf{s}'$ and $u = t-t' $. Here $d=2$,
$a$ and $c$ are non-negative scaling parameters for time and space respectively.  The smoothness parameters $\alpha$ and $\gamma$
take values in $(0,1]$, the space-time interaction parameter $\beta$ is in $[0,1]$, and $\tau \ge d/2 =1$ is, in fact, fixed at 1 following \cite{Gneiting2002}. Attractively, as $\beta$ decreases toward zero, we tend to separability in space and time.

We write the linear GP $Y(\mathbf{s},t)$ as $Y(\mathbf{s},t)=\mu_Y+\omega_Y(\mathbf{s},t)+\tilde{\varepsilon}_Y(\mathbf{s},t)$
where $\mu_Y$ is a constant mean function, $\omega_Y(\mathbf{s},t)$ is a zero mean space-time GP  with covariance function
$\sigma^2 \rho(\mathbf{h},u)$, and $\tilde{\varepsilon}(\mathbf{s},t) \stackrel{iid}{\sim} N(0,\phi_{Y}^2)$, i.e., is pure error.
It is convenient to work with the marginalized model where we integrate over
all of the $ \omega_{Y}(\mathbf{s}, t)$, see \cite{Banerjee2003}.  That is,
\begin{equation} \label{eq:modW}
Y(\mathbf{s},t)=\mu_Y+{\varepsilon}_Y(\mathbf{s},t).
\end{equation}
Then, ${\varepsilon}(\mathbf{s},t)$ is a zero mean Gaussian process with covariance function
$$\mbox{Cov}({\varepsilon}_{Y}  (\mathbf{s}_i, t_j),{\varepsilon}_{Y}  (\mathbf{s}_{i^{\prime}}, t_{j^{\prime}})) =
\sigma_Y^2\mbox{Cor}(\mathbf{h}_{i,i^{\prime}}, u_{j,j^{\prime}})+\phi_Y^2 1_{(i = i^{\prime})}1_{( j=j^{\prime})}.$$

To complete the model specification we need to specify prior distributions. We suggest the following choices.  Since
$a$ and $c$ are positive, $a$ and $c\sim G(\cdot,\cdot)$ where $G(\cdot,\cdot)$ denotes a gamma distribution.  Since $\alpha$, $\beta$, and $ \gamma$ are bounded between 0 and 1, we adopt a beta distribution ($B(\cdot,\cdot)$).
Priors for the variances and the mean direction are given the usual normal-inverse gamma form,
i.e., $\sigma^2_Y,\phi^2_Y\sim IG(\cdot,\cdot)$, where $IG(\cdot,\cdot)$ denotes the inverse gamma, and $\mu_y\sim WrapN(\cdot,\cdot)$. In the sequel, this model will be denoted by \emph{WN}.
%
%
%
%
%
\subsection{Kriging and forecasting}


We clarify prediction of the process at a new location and time, say $(\mathbf{s}_0,t_0)$, given what we have observed. We provide a full predictive distribution, extending \cite{Jona2013} who only provide a posterior mean.
Let $\mathcal{D} \subset \mathbb{R}^2\times \mathbb{Z}^+$ be the set of $n$ observed points.  Let $\mathbf{X} = \{ X(\mathbf{s},t), (\mathbf{s},t) \in \mathcal{D}  \}$   
be the vector of observed circular variables. Let  $\mathbf{Y}=\{ Y(\mathbf{s},t), (\mathbf{s},t) \in \mathcal{D}  \}$ be the associated linear ones and let $\mathbf{K}=\{ K(\mathbf{s},t), (\mathbf{s},t) \in \mathcal{D}  \}$ be the associated vector of winding numbers. The predictive distribution we seek is $g(X(\mathbf{s}_0, t_0)|\mathbf{X})$. We use usual composition sampling within MCMC to obtain samples from it. Here, again we move from the circular process to the linear one, i.e.,  a sample
from the distribution of $Y(\mathbf{s}_0, t_0)|\mathbf{X}$ can be considered as a sample from $X(\mathbf{s}_0,
t_0),K(\mathbf{s}_0, t_0)|\mathbf{X}$. If we let $\boldsymbol{\Psi}_Y$ be the vector of all parameters, we can write
\begin{equation}
	g(X(\mathbf{s}_0, t_0), K(\mathbf{s}_0, t_0)|\mathbf{X})
\end{equation}
\begin{equation}
= \sum_{\mathbf{K}\in \mathbb{Z}^n} \int_{\boldsymbol{\Psi}_Y}    g(X(\mathbf{s}_0, t_0), K(\mathbf{s}_0, t_0)|\boldsymbol{\Psi}_Y,\boldsymbol{K}, \mathbf{X}) g(\boldsymbol{\Psi}_Y,\boldsymbol{K} | \mathbf{X}) d  \boldsymbol{\Psi}_Y.
\end{equation}

So, suppose, for each posterior sample of $\mathbf{K}$ and $\boldsymbol{\Psi}_Y$ in $ \{ \mathbf{K}^*_l, \boldsymbol{\Psi}_{Y,l}^*, l=1,2,
\dots , L\}$ we generate a value from the distribution of $X(\mathbf{s}_0, t_0), K(\mathbf{s}_0, t_0)|$
$\boldsymbol{\Psi}_Y,\boldsymbol{K}, \mathbf{X} $.  Then, we will obtain the set of posterior samples $\{X^*_l(\mathbf{s}_0, t_0),$ $
K^*_l(\mathbf{s}_0, t_0) , l=1,2,\dots , L  \}$ from $X(\mathbf{s}_0, t_0), K(\mathbf{s}_0, t_0)|\mathbf{X}$. If, we  retain  the set
$\{ X^*_l(\mathbf{s}_0, t_0) , l=1,2,\dots , L  \}$, we will have samples from the desired predictive distribution.

Therefore, we need  to sample from the distribution of  $X(\mathbf{s}_0, t_0), K(\mathbf{s}_0,  t_0)|$
$\boldsymbol{\Psi}_Y,\boldsymbol{K}, \mathbf{X}$ or equivalently $Y(\mathbf{s}_0, t_0)|\mathbf{Y}, \boldsymbol{\Psi}_Y $. Let $\mathbf{1}_m$ be the $m \times 1$ vector of 1s,  let $\boldsymbol{C}_\mathbf{Y}$ be the correlation matrix of $\boldsymbol{Y}$, and let $\boldsymbol{C}_{\mathbf{Y},Y(\mathbf{s}_0,t_0)}$ be the correlation vector between $\mathbf{Y}$ and $Y(\mathbf{s}_0,t_0)$.  Then, the joint distribution of $Y(\mathbf{s}_0,t_0), \mathbf{Y}| \boldsymbol{\Psi}_Y$ is
\begin{equation}
	\left(
	\begin{array}{c}
		Y(\mathbf{s}_0,t_0) \\
		\mathbf{Y}
	\end{array}
	\right)| \boldsymbol{\Psi}_Y \sim N
	\left(
	\left(
	\begin{array}{c}
		\mu_Y\\
		\mu_Y \mathbf{1}_n
	\end{array}
	\right),
	\sigma_Y^2\left(
	\begin{array}{cc}
		1&   \boldsymbol{C}_{\mathbf{Y},Y(\mathbf{s}_0,t_0)}^{\prime}\\
		\boldsymbol{C}_{\mathbf{Y},Y(\mathbf{s}_0,t_0)}& \mathbf{C}_\mathbf{Y}
	\end{array}
	\right)+ \phi_Y^2 \mathbf{I}_{n+1}
	\right).
\end{equation}
As a result, the conditional distribution of $Y(\mathbf{s}_0,t_0)| \mathbf{Y}, \boldsymbol{\Psi}_Y   $ is Gaussian with mean
$$\mbox{M}_{Y(\mathbf{s}_0,t_0)} = \mu_Y  +    \sigma_Y^2\boldsymbol{C}_{\mathbf{Y},Y(\mathbf{s}_0,t_0)}^{\prime}  \left(\sigma_Y^2\mathbf{C}_\mathbf{Y}   + \phi_Y^2 \mathbf{I}_n \right)^{-1}\left( \mathbf{Y}-\mu_Y \mathbf{1}_n
\right)
$$
and variance
$$
\mbox{V}_{Y(\mathbf{s}_0,t_0)} = \sigma_Y^2+\phi_Y^2-\sigma_Y^2\boldsymbol{C}_{\mathbf{Y},Y(\mathbf{s}_0,t_0)}^{\prime}  \left(\sigma_Y^2\mathbf{C}_\mathbf{Y}   + \phi_Y^2 \mathbf{I}_n \right)^{-1} \sigma_Y^2\boldsymbol{C}_{\mathbf{Y},Y(\mathbf{s}_0,t_0)}.
$$

Finally, suppose, for each posterior sample, we simulate  $Y^*_l(\mathbf{s}_0,t_0) $ from $N(\mbox{M}_{Y(\mathbf{s}_0,t_0),l}^*,
\mbox{V}_{Y(\mathbf{s}_0,t_0),l}^*)$,
 where $\mbox{M}_{Y(\mathbf{s}_0,t_0),l}^*$ and $ \mbox{V}_{Y(\mathbf{s}_0,t_0),l}^*$ are
$\mbox{M}_{Y(\mathbf{s}_0,t_0)}$ and $ \mbox{V}_{Y(\mathbf{s}_0,t_0)}$  computed with the $l^{th}$ sample.  then, $
X^*_l(\mathbf{s}_0,t_0) = Y^*_l(\mathbf{s}_0,t_0) \hbox{    mod } 2 \pi $ is a posterior sample from the predictive
distribution.

\section{The spatio-temporal projected normal process} \label{sec:projM}

Let  $ (Z_1, Z_2)$ be  a bivariate vector normally distributed with mean $\boldsymbol{\mu}_Z = (\mu_{Z_1}, \mu_{Z_2})$ and
covariance matrix
\begin{equation}
	\tilde{\mathbf{V}}=\left(
	\begin{array}{cc}
		\sigma_{Z_1}^2 & \sigma_{Z_1}\sigma_{Z_2} \rho_z\\
		\sigma_{Z_1}\sigma_{Z_2} \rho_z & \sigma_{Z_2}^2
	\end{array}
	\right).
\end{equation}
The vector $\mathbf{Z}$ is mapped into  an angular variable $\Theta$ by the transformation $\Theta = {\atan}^*
({Z_2}/{Z_1})$, where the function $\atan^*(S/C)$ is defined as $\atan(S/C)$ if $C > 0$ and $S \geq 0$, $\pi / 2$  if $C=0$ and  $S > 0$, $\atan(S/C)+\pi$ if $C < 0$, $\atan(S/C)+2\pi$ if $C \geq 0$ and $S < 0$, undefined if $C=S=0$. $\Theta$ is referred to as a projected normal random variable  \citep[][p. 52]{mardia72} with parameters $\boldsymbol{\mu}_{Z}$
and $\tilde{\mathbf{V}}$. \cite{Wang2013}
note that the distribution
of $\Theta$ does not change if we multiply $(Z_1,Z_2)$ by a positive constant, so, following their
lead, to identify the distribution  we set $\sigma_{Z_2}^2=1$ and the covariance matrix becomes
\begin{equation}
	\mathbf{V} =  \left(
	\begin{array}{cc}
		\sigma_{Z_1}^2 & \sigma_{Z_1} \rho_z\\
		\sigma_{Z_1} \rho_z &1
	\end{array}
	\right).
\end{equation}
Again, it is convenient to introduce a latent variable.  Here, it is $R = ||\mathbf{Z}||$, obtaining the joint density of $(\Theta, R)$:
\begin{equation}
	(2 \pi)^{-1} |\mathbf{V}|^{{1/2}}\exp \left( - \frac{(r(\cos \theta, \sin \theta)^{\prime}-\boldsymbol{\mu}_Z)^{\prime}\mathbf{V}^{-1}(r(\cos \theta, \sin \theta)^{\prime}-\boldsymbol{\mu}_Z)}{2} \right) r.
\end{equation}
We can move back and forth between the linear variables and the pair  $(\Theta, R)$ using   the transformation  $Z_1 = R \cos \Theta$, $Z_2 = R \sin \Theta$ and the  equation $\Theta = {\atan}^*
({Z_2}/{Z_1})$.

Consider a bivariate spatio-temporal process $\mathbf{Z}(\mathbf{s},t) =
(Z_1(\mathbf{s},t),Z_2(\mathbf{s},t))$ with constant mean $\boldsymbol{\mu}_Z$ and cross covariance function
$\mbox{C}\left(\mathbf{Z}   (\mathbf{s}_i,t_j), \mathbf{Z}(\mathbf{s}_{i^{\prime}},t_{j^{\prime}})\right) =\mbox{Cor}(
\mathbf{s}_i- \mathbf{s}_{i^{ \prime }},t_j-t_{j^{\prime}}) \mathbf{V} $ where $\mbox{Cor}(\cdot,\cdot)$ is a given space-time
correlation function and $\mathbf{V}$ is as above. Then the circular process $\Theta(\mathbf{s},t)$ induced by
$\mathbf{Z}(\mathbf{s},t) $ with the ${\atan}^*$ transformation is a projected Gaussian process with mean
$\boldsymbol{\mu}_Z$ and covariance function induced by $\mbox{C}\left(\mathbf{Z}   (\mathbf{s}_i,t_j),
\mathbf{Z}(\mathbf{s}_{i^{\prime}},t_{j^{\prime}})\right)$. More details on the properties of the process  can be found in
\citet{wang2014}. Now, latent $R(\mathbf{s},t)$'s are introduced to facilitate  model fitting.

\subsection{Model specification and model fitting}\label{sec:predP}
 We define the bivariate linear process as
\begin{equation}  \label{eq:projz}
	\begin{array}{cc}
		Z_{\ell}(\mathbf{s}, t) = \mu_{Z_\ell}+ \omega_{Z_\ell}(\mathbf{s}, t)+\tilde{\varepsilon}_{Z_\ell}(\mathbf{s}, t), \quad \ell=1,2, \\
	\end{array}
\end{equation}
where $\boldsymbol{\mu}_{Z} = (\mu_{Z_1},\mu_{Z_2})^{\prime}$  is the mean level,  $\boldsymbol{\omega}_{Z}(\mathbf{s}, t) =
(\omega_{Z_1}(\mathbf{s}, t), \omega_{Z_2}(\mathbf{s}, t))^{\prime}$ is a bivariate Gaussian process with zero mean and
covariance $\mbox{Cov}(\boldsymbol{\omega}_{Z}(\mathbf{s}_i,t_j), $  $
\boldsymbol{\omega}_{Z}(\mathbf{s}_{i^{\prime}},t_{j^{\prime}}))    = \mbox{Cor}( \mathbf{h}_{i,i^{\prime}},u_{j,j^{\prime}})
\mathbf{V}$ where  $\mbox{Cor}(  \mathbf{h}_{i,i^{\prime}},u_{j,j^{\prime}}) $ is defined in \eqref{eq:cov1}.  Finally,
$\tilde{\boldsymbol{\varepsilon}}_{Z}  (\mathbf{s}, t)=(\tilde{\varepsilon}_{Z_1}(\mathbf{s},
t),\tilde{\varepsilon}_{Z_2}(\mathbf{s}, t)) $ is   bivariate pure error with zero mean, independent components, and variance $\phi_{{Z}}^2$.
%
Marginalizing over the $\omega$ process in \eqref{eq:projz} yields
\begin{equation}\label{eq:modP}
Z_{\ell}(\mathbf{s}, t) = \mu_{Z_\ell}+ \varepsilon_{Z_\ell}(\mathbf{s}, t), \quad \ell=1,2,
\end{equation}
%
where $\boldsymbol{\varepsilon}_{Z}  (\mathbf{s}, t)$ is a mean zero bivariate Gaussian process with  covariance function
$\mbox{Cov}(\boldsymbol{\varepsilon}_{Z}  (\mathbf{s}_i, t_j),$ $\boldsymbol{\varepsilon}_{Z}  (\mathbf{s}_{i^{\prime}},
t_{j^{\prime}}) $ $ = \mbox{Cor}(\mathbf{h}_{i,i^{\prime}},u_{j,j^{\prime}})
 \mathbf{V}+ \phi_Z^2\mathbf{I}_21_{(i = i^{\prime})}1_{(j = j^{\prime})}
	$.

 $\Theta(\mathbf{s},t)=\atan^* ({Z_2(\mathbf{s},t)} / {Z_1(\mathbf{s},t)})$ is a circular process
and, as in the WN setting,  correlation
between the circular variables is induced by the Gneiting spatio-temporal correlation function.
To specify the prior distributions for $\mu_{Z_1}$, $\mu_{Z_2}$,  $\sigma_{Z_1}^2$ and $\phi_{Z}^2$,  we adopt the customary normal-inverse gamma specification. That is,  $\mu_{Z_1},\mu_{Z_2} \sim N(\cdot, \cdot)$,  $\sigma_{Z_1}^2,\phi_{Z}^2 \sim IG(\cdot,\cdot)$ while, since $\rho_Z$ is a correlation parameter, we adopt a truncated normal:  $\rho_Z \sim N(\cdot, \cdot)I(-1,1)$.
In the sequel, this model will be denoted by \emph{PN}.

We seek the predictive distribution at an  unobserved location and time, $(\mathbf{s}_0,t_0)$. Let $\boldsymbol{\Theta}$ be the
vector of  observed circular values and $\mathbf{Z}=\{ \mathbf{Z}(\mathbf{s},t), (\mathbf{s},t) \in \mathcal{D}  \}$ be the
associated linear ones. Let  $\mathbf{Z}(\mathbf{s}_0,t_0) = (Z_1(\mathbf{s}_0,t_0),Z_2(\mathbf{s}_0,t_0))^{\prime}$,
$\mathbf{R} = \{ R(\mathbf{s},t), (\mathbf{s},t) \in \mathcal{D} \}$ and let $\boldsymbol{\Psi}_{Z}$ be all the
parameters of the projected model.

Specifically, the predictive distribution we seek is  $\Theta (\mathbf{s}_0,t_0)|
\boldsymbol{\Theta}$.
If we sample from the distribution of $\mathbf{Z} (\mathbf{s}_0,t_0)| \boldsymbol{\Theta}$ then
$\Theta(\mathbf{s}_0,t_0)$ $=\atan^* ({Z_2(\mathbf{s}_0,t_0)} / {Z_1(\mathbf{s}_0,t_0)})$ is a sample from the desired predictive
distribution. We have that
\begin{equation}
	g(\mathbf{Z} (\mathbf{s}_0,t_0)| \boldsymbol{\Theta}) = \int_{\mathbf{R}} \int_{\boldsymbol{\Psi}_Z} g(\mathbf{Z} (\mathbf{s}_0,t_0)|\boldsymbol{\Psi}_Z,\mathbf{R}, \boldsymbol{\Theta}) g(\boldsymbol{\Psi}_Z,\mathbf{R}| \boldsymbol{\Theta}) d \boldsymbol{\Psi}_Z d \mathbf{R}.
\end{equation}
So, we need to obtain $ g(\mathbf{Z} (\mathbf{s}_0,t_0)|\boldsymbol{\Psi}_Z,\mathbf{R}, \boldsymbol{\Theta})$ and be able to
sample from it. We start from the joint distribution of $\mathbf{Z}(\mathbf{s}_0,t_0) , \mathbf{Z}| \boldsymbol{\Psi}_Z$:
\begin{equation}
	\left(
	\begin{array}{c}
		\mathbf{Z}(\mathbf{s}_0,t_0) \\
		\mathbf{Z}
	\end{array}
	\right)| \boldsymbol{\Psi}_Z
	\end{equation}
	\begin{equation}
	\sim N
	\left(
	\left(
	\begin{array}{c}
		\boldsymbol{\mu}_Z\\
		\mathbf{1}_n \otimes \boldsymbol{\mu}_{Z}
	\end{array}
	\right),
	\left(
	\begin{array}{cc}
		1 &\boldsymbol{C}_{\mathbf{Z},\mathbf{Z}(\mathbf{s}_0,t_0)}^{\prime}\\
		\boldsymbol{C}_{\mathbf{Z},\mathbf{Z}(\mathbf{s}_0,t_0)}& \mathbf{C}_\mathbf{Z}
	\end{array}
	\right) \otimes \mathbf{V} +\phi_Z^2 \mathbf{I}_{2n+2}
	\right),
\end{equation}
where  $\mathbf{C}_\mathbf{Z}$ and $\boldsymbol{C}_{\mathbf{Z},\mathbf{Z}(\mathbf{s}_0,t_0)}$ are the analogous of   $\mathbf{C}_\mathbf{Y}$ and $\boldsymbol{C}_{\mathbf{Y},Y(\mathbf{s}_0,t_0)}$ for the process $\mathbf{Z}(\mathbf{s},t)$.
The conditional distribution of $\mathbf{Z}(\mathbf{s}_0,t_0) | \mathbf{Z}, \boldsymbol{\Psi}_Z$ (equivalently $\mathbf{Z}(\mathbf{s}_0,t_0) | \boldsymbol{\Theta}, \mathbf{R}, \boldsymbol{\Psi}_Z$) is bivariate  normal with mean
$$
\mbox{M}_{\mathbf{Z}(\mathbf{s}_0,t_0)} = \boldsymbol{\mu}_Z  + \boldsymbol{C}_{\mathbf{Z},\mathbf{Z}(\mathbf{s}_0,t_0)}^{\prime}\otimes \mathbf{V} \left(  \mathbf{C}_\mathbf{Z} \otimes \mathbf{V} + \phi_Z^2 \mathbf{I}_{2n}   \right)^{-1}(\mathbf{Z}- \mathbf{1}_n \otimes \boldsymbol{\mu}_{Z})
$$
and variance
$$
\mbox{V}_{\mathbf{Z}(\mathbf{s}_0,t_0)} = \mathbf{V}- \boldsymbol{C}_{\mathbf{Z},\mathbf{Z}(\mathbf{s}_0,t_0)}^{\prime}\otimes \mathbf{V} \left(  \mathbf{C}_\mathbf{Z} \otimes \mathbf{V} + \phi_Z^2 \mathbf{I}_{2n}   \right)^{-1} \boldsymbol{C}_{\mathbf{Z},\mathbf{Z}(\mathbf{s}_0,t_0)}\otimes \mathbf{V}.
$$
Using the posterior samples $\{\mathbf{R}_l^*, \boldsymbol{\Psi}_{Z,l}^* , l=1,2,\dots ,L\}$ we can collect samples of
$\Theta_l^*(\mathbf{s}_0,t_0)$ from its posterior predictive distribution.
\section{Simulated examples} \label{sec:sim}

\begin{figure}[t!]
	\centering
	\subfloat[WN]{\includegraphics[scale=0.345]{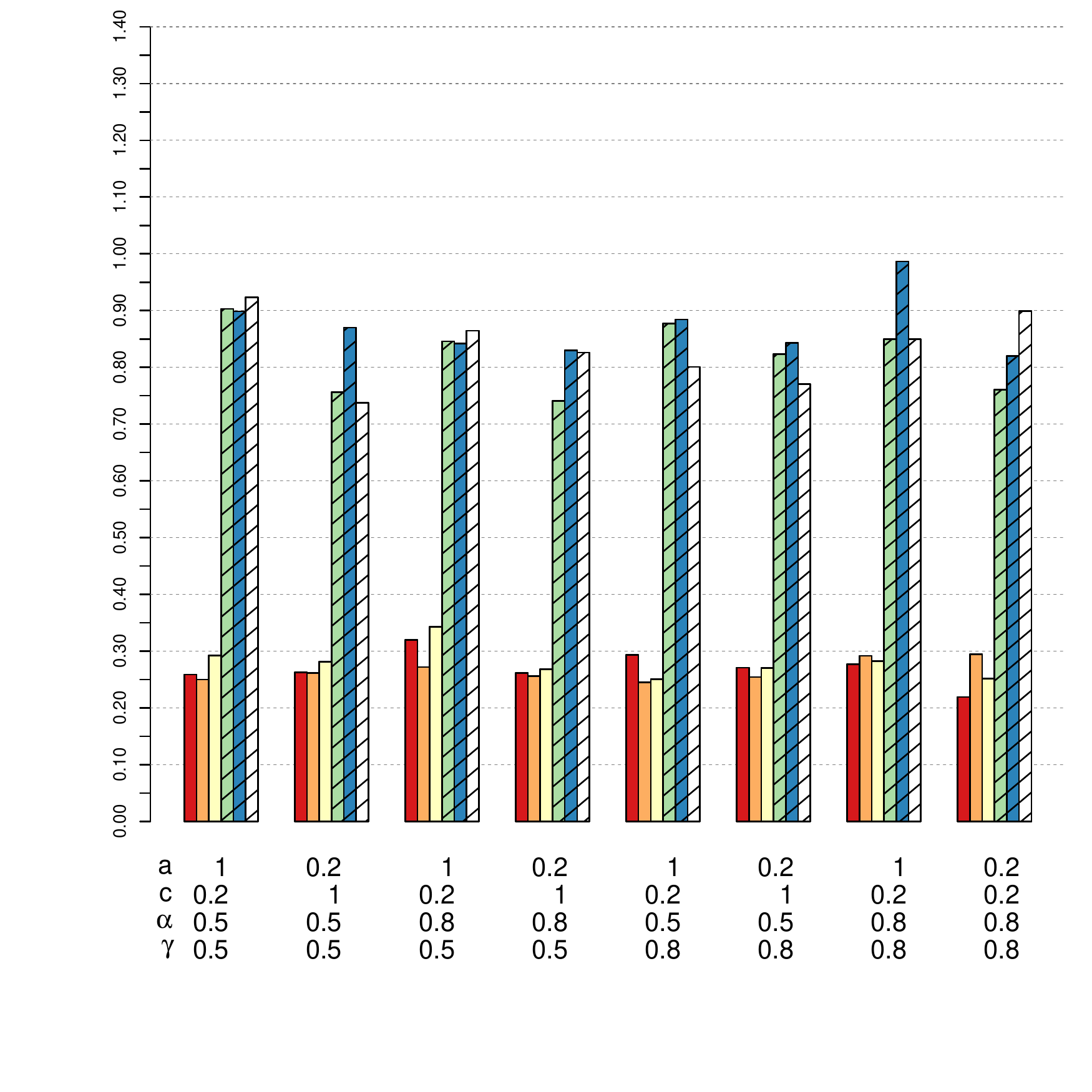}}
	\subfloat[PN]{\includegraphics[scale=0.345]{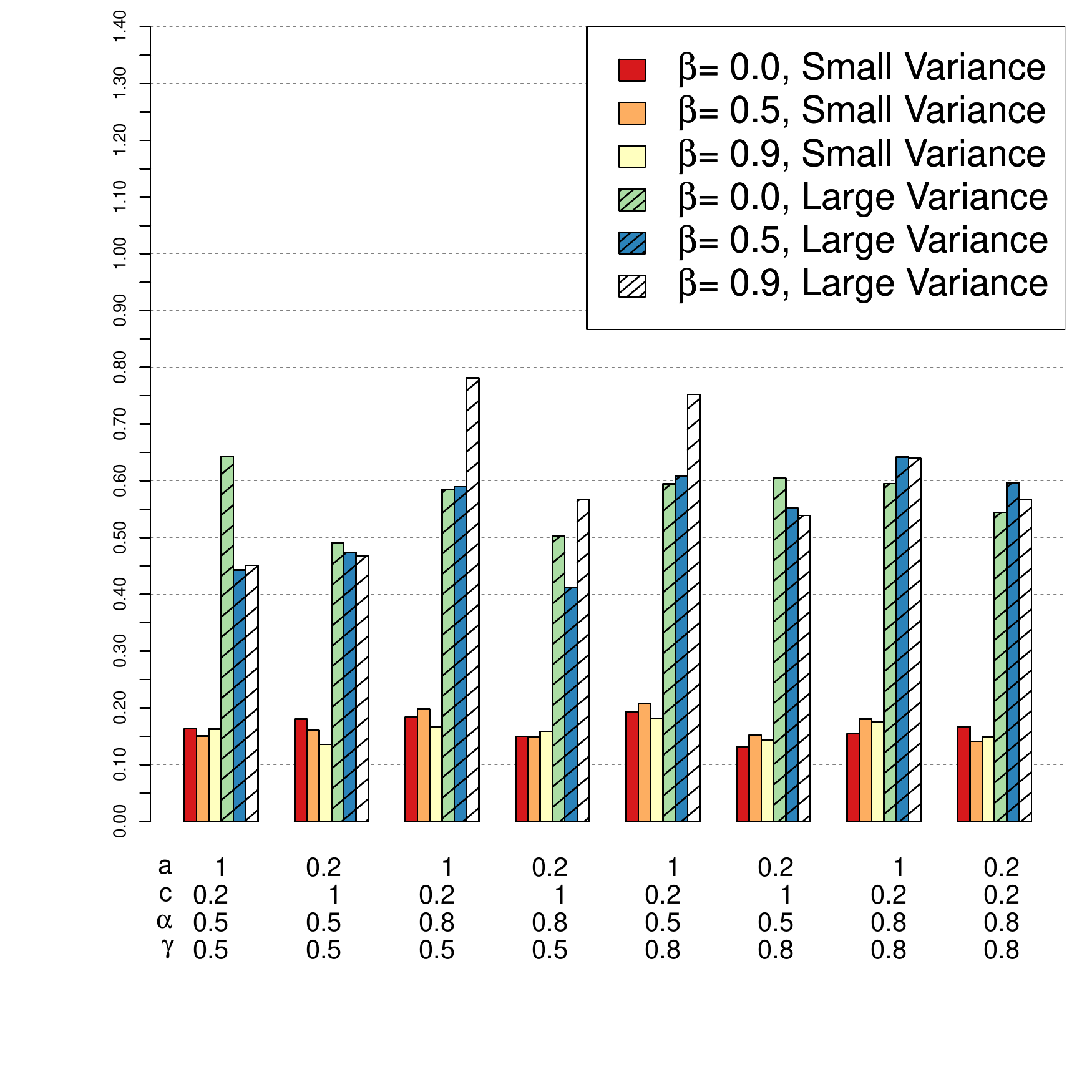}}
	\caption{Simulation study: CRPS comparing performances of the two proposed models} \label{fig:CRPS1}
\end{figure}

The Gneiting correlation function \eqref{eq:cov1} has not been widely investigated within a Bayesian framework. The aim of this
simulation study is essentially to provide a proof of concept.  If space-time dependence, captured through the Gneiting correlation
function, is driving an observed spatio-temporal circular dataset, can we learn about this dependence and can we demonstrate
improved predictive performance by incorporating it in our modeling?  We explore several different choices of parameters in
\eqref{eq:cov1}.

For each proposed model we simulated 48 datasets with $n=240$ (20 locations and 12 time points) with  spatial coordinates
uniformly generated in $[0,10]\times[0,10]$.  24 datasets for the WN  model were simulated from all possible combinations of
$(a,c)= \{(1,0.2), (0.2,1) \}$, $\beta=\{0,0.5,1\}$, $\alpha=\{ 0.5,0.8  \}$, $\gamma=\{ 0.5,0.8  \}$ and $(\mu_{Y},\sigma_Y^2,
\phi_y^2) = (\pi,0.1, 0.01)$.  In the other 24 datasets we used the same combinations of correlation parameters but with
$(\mu_{Y},\sigma_Y^2, \phi_y^2) = (\pi,1, 0.1)$. The datasets  cover a wide range of situations in terms of spatio-temporal
correlation: strong spatial correlation with weak temporal correlation $((a,c)=(1,0.2))$,   weak spatial correlation with strong
temporal correlation $((a,c)=(0.2,1))$, fully separable spatio-temporal correlation ($\beta=0$), non-separable ($\beta=\{0.5,0.9\}$) and two levels for the smoothing parameters.
The difference between the two collections of 24 datasets is that the first 24 have smaller circular variance than the remaining ones, where the  circular
variance was computed as one minus the mean resultant length divided by the sample size \citep[p.
	15]{Jammalamadaka2001}.

The projected normal datasets were built according to the same rationale adopted for the wrapped normal, i.e. we built 24 datasets with
small circular  variance  and 24 datasets with large circular variance. We simulated from unimodal projected distributions
adopting the following sets of parameters:
\begin{itemize}
	\item all possible combinations of $(a,c)= \{(1,0.2), (0.2,1) \}$, $\beta=\{0,0.5,1\}$, $\alpha=\{ 0.5,0.8  \}$, $\gamma=\{
	0.5,0.8  \}$  with $(\mu_{Z_1}, \mu_{Z_2}, \sigma_{Z_1}^2, \rho_Z,\phi_Z^2)= (2.5,2.5,1,0,0.01)  $ which yields a
	circular variance  close to the   WN examples with $\sigma_Y^2=0.1$.
	\item  all possible combinations of $(a,c)= \{(1,0.2), (0.2,1) \}$, $\beta=\{0,0.5,1\}$, $\alpha=\{ 0.5,0.8  \}$, $\gamma=\{
	0.5,0.8  \}$ with $(\mu_{Z_1}, \mu_{Z_2}, \sigma_{Z_1}^2, \rho_Z ,	\phi_Z^2)= (0.85,0.85,1,0,0.1)  $ which, again,
	yields a circular variance  close to the WN examples with $\sigma_Y^2=1$.
\end{itemize}
The parameters for the prior distributions were chosen so that the priors were centered on the ``true'' values used to
simulate each dataset:
\begin{itemize}
	\item correlation parameters:
	$a=0.2 \Rightarrow a \sim G(2,5)$,
	$a=1 \Rightarrow a \sim G(5,4)$,
	$c=0.2 \Rightarrow c \sim G(2,5)$,
	$c=1 \Rightarrow c \sim G(5,4)$,
	$\alpha=0.5 \Rightarrow \alpha \sim B(5,5)$,
	$\alpha=0.8 \Rightarrow \alpha \sim B(6,1.5)$,
	$\beta=0 \Rightarrow \beta \sim B(1,4)$,
	$\beta=0.5 \Rightarrow \beta \sim B(5,5)$,
	$\beta=0.9 \Rightarrow \beta \sim B(6,1.5)$,
	$\gamma=0.5 \Rightarrow \gamma \sim B(5,5)$,
	$\gamma=0.8 \Rightarrow \gamma \sim B(6,1.5)$;
	\item parameters of the WN model: $\mu_Y=5 \Rightarrow  \mu_Y \sim WN(\pi,5)$,
	$\sigma_{Y}^2 =0.1 \Rightarrow \sigma_{Y}^2 \sim IG(4.5, 0.55)$,
	$\sigma_{Y}^2 =1 \Rightarrow \sigma_{Y}^2 \sim IG(2.01, 4.01)$,
	$\phi_{Y}^2 =0.01 \Rightarrow \phi_{Y}^2 \sim IG(2.001, 0.03)$,
	$\phi_{Y}^2 =0.1 \Rightarrow \phi_{Y}^2 \sim IG(4.5, 0.55)$;
	\item parameters of the PN model:
	$\mu_{Z_1} = 2.5 \Rightarrow \mu_{Z_1} \sim N(2.5,5)$,
	$\mu_{Z_1} = 0.85 \Rightarrow \mu_{Z_1} \sim N(0.85,5)$
	$\mu_{Z_2} = 2.5 \Rightarrow \mu_{Z_2} \sim N(2.5,5)$,
	$\mu_{Z_2} = 0.85 \Rightarrow \mu_{Z_2} \sim N(0.85,5)$,
	$\sigma_{Z_1}^2 = 1 \Rightarrow \sigma_{Z_1}^2 \sim IG(2.01,4.01)$,
	$\rho_{Z} = 0   \Rightarrow  \rho_{Z}\sim  N(0,1)I(-1,1)$,
	$\phi_{Z}^2 =0.01 \Rightarrow \phi_{Z}^2 \sim IG(2.001, 0.03)$,
	$\phi_{Z}^2 =0.1 \Rightarrow \phi_{Z}^2 \sim IG(4.5, 0.55)$.
\end{itemize}

Among the 240 simulated observations in each dataset, 170 points, chosen between the first  and tenth time points, were used for estimation
and the remaining 70 points were set aside for validation purposes. The predictive performance was evaluated using two criteria.
We computed an \emph{average prediction error} (APE), defined as the average circular distance
between a validation
dataset and model predicted values, where we adopted as circular distance $d(\alpha,\beta)=1-\cos(\alpha-\beta)$ \cite[p.15]{Jammalamadaka2001}. In particular, suppose the validation set has $n^*$ observations, the APE for the models based on the
wrapped normal is $ \frac{1}{n^*} \sum_{(\mathbf{s}_0,t_0)}d(\mu(\mathbf{s}_0,t_0| \mathbf{X}),
x(\mathbf{s}_0, t_0))$ and $ \frac{1}{n^*} \sum_{(\mathbf{s}_0,t_0)} d(\mu(\mathbf{s}_0,t_0|
\boldsymbol{\Theta}), \theta(\mathbf{s}_0, t_0))$ for the projected normal ones. Here, $x(\mathbf{s}_0, t_0)$ and $	
\theta(\mathbf{s}_0, t_0)$ are the realizations of the processes at $(\mathbf{s}_0,t_0)$ and $\mu(\mathbf{s}_0,t_0| \mathbf{X})$ and $\mu(\mathbf{s}_0,t_0| \boldsymbol{\Theta})$ are the posterior mean directions.

We also computed the \emph{continuous ranked probability score} (CRPS) for circular variables as defined in \cite{grimit2006}:
\begin{equation} \label{eq:crps}
	CRPS(F, \delta)= E(d(\Delta,\delta))-\frac{1}{2}E(d(\Delta,\Delta^*)),
\end{equation}
where $F$ is a predictive distribution, $\delta$ is a holdout value, and $\Delta$  and $\Delta^*$ are independent copies of a
circular variable with distribution $F$.
In this form, small values of CRPS are preferred. 

For both models we do not know $F$ in
closed form but we can compute a Monte Carlo approximation of  \eqref{eq:crps}. For the wrapped model, the CRPS for a held-out
space-time point $(\mathbf{s}_0,t_0)$ is
\begin{equation}
 \frac{1}{L}\sum_{l=1}^L d(x_l^*(\mathbf{s}_0,t_0),x(\mathbf{s}_0,t_0))-\frac{1}{2L^2}\sum_{l=1}^L\sum_{j=1}^L d(x_l^*(\mathbf{s}_0,t_0),x_j^*(\mathbf{s}_0,t_0)) \label{eq:crps1}
\end{equation}
and for the projected model it is
\begin{equation}
\frac{1}{L}\sum_{l=1}^L d(	\theta_l^*(\mathbf{s}_0,t_0),\theta(\mathbf{s}_0,t_0))-\frac{1}{2L^2}\sum_{l=1}^L\sum_{j=1}^L  d (\theta_l^*(\mathbf{s}_0,t_0),\theta_j^*(\mathbf{s}_0,t_0)).\label{eq:crps2}
\end{equation}

For each  of the 48 simulated datasets, the values of the mean CRPS under the two models, computed over the set of points used for  model validation,
 are shown in Figure \ref{fig:CRPS1}. For both models we see that the CRPS depends heavily on the variance
of the process but seems unaffected by changes in the other parameters.

A potentially important difference between the two models is the computational time required to fit them. The WN model is
computationally more efficient than the PN model; the main issue is computational complexity (see Supplementary Online Material, Section S1).
The PN requires, at each MCMC iteration, roughly 8 times as many operations as the WN to be fitted. If computational time is a relevant issue, then the WN may be more attractive.


\section{Real Data} \label{sec:real}

\begin{figure}[t!]
	\centering 
	\subfloat[]{\includegraphics[scale=0.34]{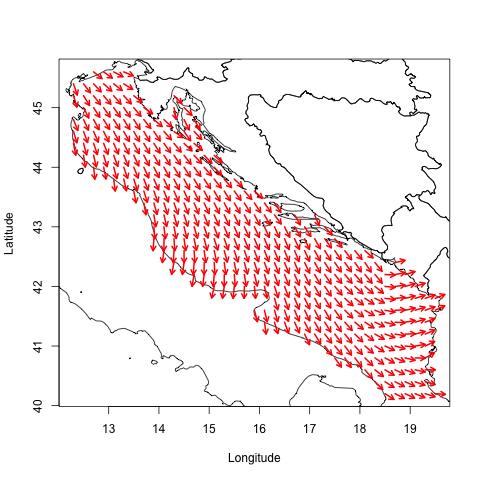}}
	\subfloat[ ]{\includegraphics[scale=0.34]{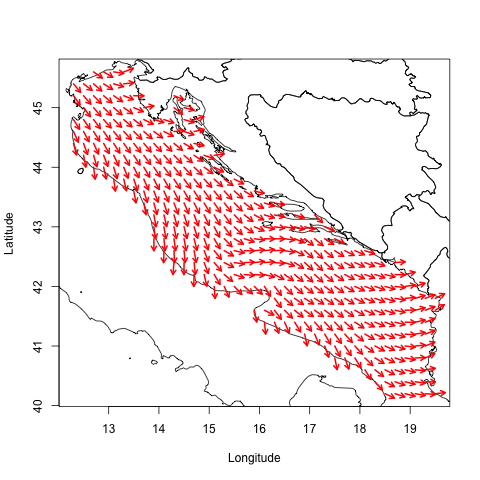}} \\
	\subfloat[ ]{\includegraphics[scale=0.34]{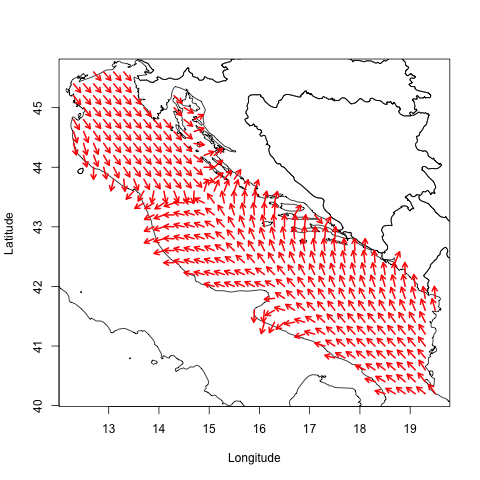}}
	\subfloat[]{\includegraphics[scale=0.34]{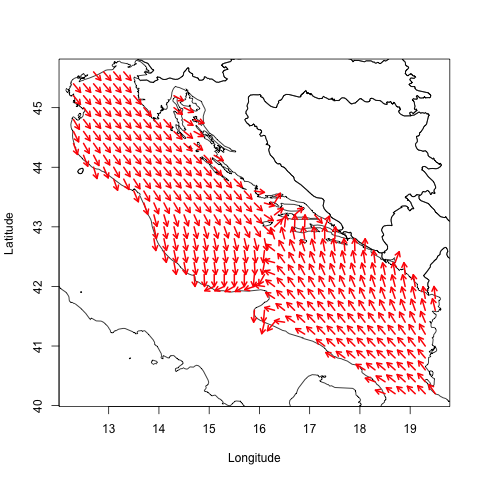}}
	\caption{Time windows for different sea states used for validation. The four panels represent the observed wave direction over the entire area at: (a) 12:00 on 5/5/2010 (storm); (b) 00:00 on 6/5/2010 (transition between storm and calm); (c) 00:00 on 7/5/2010 (calm); (d) 12:00 on 7/5/2010 (one-step prediction, calm).}   \label{fig:sec}
\end{figure}


We model  wave directions obtained as outputs from a deterministic computer model implemented by Istituto
Superiore per la Protezione e la Ricerca Ambientale (ISPRA). The computer model starts from a wind forecast model predicting the
surface wind   over the entire Mediterranean. The hourly evolution of sea wave spectra is obtained by solving energy transport
equations using the wind forecast as input. Wave spectra are locally modified using a source function describing the wind
energy, the energy redistribution due to nonlinear wave interactions, and energy dissipation due to wave fracture. The model
produces estimates every hour on a grid with 10$\times$10 km cells \citep{speranza2004, MET:MET34}. The ISPRA dataset has
forecasts for a total of 4941 grid points over the Italian Mediterranean. Over the Adriatic Sea area, there are 1494 points.
%
%
%
%

Our aim is to  compare the performance of the WN and PN models. From a phenomenological perspective, the PN model is arguably
the more natural choice since we are not wrapping a linear scale to obtain the directions.  However, the WN model does provide  a suitable model and, as suggested  above, it may be  attractive in terms of computational efficiency and interpretability of  parameters.
In the selected dataset, the
three sea states, {\em calm, transition} and \emph{storm} are present. The sea state is defined through the wave height
(which is also supplied by the computer model output): when this height is below 1 meter, we have \emph{calm}, when it is between 1 and
2 meters we have \emph{transition}  (between calm and storm) and when it is greater than 2 meters we have a \emph{storm}. Wave
directions vary more in calm than in storm.
Here, we seek to learn about  the spatio-temporal structure of the data relying only on the specification of the correlation function.  We will use the information given by the  wave heights in the models proposed
in Section \ref{sec:real2}.

%
We fitted the model using 100 spatial points $\times$ 10 time points six hours apart (1000 observations in total) in order to have a dataset including all sea states. Notice that spatial distances are evaluated in kilometres. Then, we developed four validation datasets,
each with 350 spatial points and 1 time point.  Specifically, we have one dataset for each sea state plus one for a one-step forward prediction.  Finally, we used the model fitted
over the 1000 points to predict  each validation dataset.
Three of the datasets are inside the time window used for model estimation, one in calm sea, one in transition and one during a storm.
The fourth validation set is at 12:00 on May 7, 2010,  6 hours after the last time used for model fitting.  The observed circular
process in each of these four time windows can be seen in Figure \ref{fig:sec}. For each time window and model we computed the mean CRPS
and APE, see Table \ref{tab:CRPS}. Furthermore, we computed the mean CRPS and APE over the 4 time windows.
\begin{table}[t!]
	\centering
	\caption{Real data example:  CRPS and APE  for the WN and PN models computed on each validation dataset.} \label{tab:CRPS}
	\begin{tabular}{c|c|c|c}
		 \hline \hline
		  & & WN & PN \\ \hline \hline
		  Average & CRPS &0.655 & 0.629\\
		  	& APE & 0.437 &0.421\\\hline
		  Calm & CRPS &1.450&1.398\\
		  & APE &0.995 &0.973\\\hline
		  Transition & CRPS &0.082&0.074\\
		  & APE &0.033&0.028\\\hline
		  Storm & CRPS &0.063&0.042\\
		  & APE &0.026&0.009\\\hline
		  One-step prediction & CRPS &1.024&1.001\\
		  & APE &0.693&0.674\\ \hline \hline
	\end{tabular}
\end{table}

%

Following our discussion in Sections \ref{sec:WNmodfit} and \ref{sec:projM} we used the following priors:  $a\sim G(1.5,1)$, $c\sim G(1.5,1)$, $\alpha\sim B(2,2.5)$, $\beta\sim B(1.1,2)$, $\gamma\sim
B(2,2.5)$, $\sigma_Y^2\sim IG(2,2)$, $\phi_Y^2\sim IG(1,0.25)$, $\mu_Y\sim WrapN(\pi,10)$,  $\mu_{Z_1} \sim N(0,10)$, $\mu_{Z_2}
\sim N(0,10)$, $\rho_{Z} \sim N(0,5)I(-1,1)$, $\sigma_Z^2\sim IG(2,2)$ and $\phi_Z^2 \sim IG(1, 0.25)$.  Notice that all distributions are weakly informative. Also, the prior for $\beta$ is centered near 0.1, i.e. close to the  separable model. Decay parameters in space and time are related to the minimum and maximum distances in space and time, chosen to ensure that they concentrate the probability mass over such intervals.


As we expected, the predictive capability of the two models, in terms of both CRPS and APE, is poorest in a calm state, the variance
being larger than in other states.  On the other hand, it is very accurate during a storm or a transition for both models as we
can see in Table \ref{tab:CRPS}. The PN always performs  better that the WN. The largest difference between the APE values of the
two models (0.022) is observed during the calm sea time window.

\begin{table}[t!]
	\centering
	\caption{Real data example: mean point estimate (PE) and 95\% credible interval (CI)  for the correlation parameters for the WN and PN models } \label{tab:cp1}
	\begin{tabular}{c|c|c|c}
		\hline\hline
		&& WN & PN \\ \hline\hline
		$a$ & PE & 	0.076 & 0.009\\	
		& (CI) &  (0.019,0.200)&(0.005,0.019)\\\hline
		$c$ & PE &  $3.2 \times 10^{-4}$& 1.4$\times 10^{-4}$\\
		& (CI) & (1.3$\times 10^{-4}$,7.1$\times 10^{-4}$)& (7.0$\times 10^{-4}$,2.9$\times 10^{-4}$)\\\hline
		$\alpha$ & PE   & 0.495 & 0.693 \\
		& (CI) & (0.288,0.744) &  (0.562,0.819) \\\hline
		$\beta$ & PE & 0.592 & 0.430 \\
		& (CI) & (0.158,0.915) &(0.101,0.774) \\\hline
		$\gamma$ & (PE) & 0.797&0.872\\
		& (CI) & (0.697 0.897)&(0.779,0.939) \\\hline\hline
	\end{tabular}
\end{table}

In Table \ref{tab:cp1}  we give credible intervals and posterior mean estimates for the value of the
parameters of the correlation function.  For both models nonseparable correlation structure is strongly supported.
The point  estimates of the spatial $(c)$ and temporal $(a)$ decay  are smaller in the PN model.
Notice that  data are bimodal whenever the wave directions look like those in Figure \ref{fig:sec} (c) and (d), i.e., when over a large region at a given time a storm is rotating or two different weather systems are meeting. Then, scalar statistics, such as the overall mean direction or the overall concentration, may not be informative regarding this behaviour.

In the Supplementary Online Material,  we provide the parameter estimates for the wrapped and projected distributions with associated 95\%
credible intervals (Table S1). Since $\mu_Y$ is defined on a circular domain (recall that
the prior on $\mu_Y$ is  $WrapN(\cdot, \cdot)$), following \cite{Jona2013}, we can compute a 95\%
credible interval as the arc that contains the central 95\% of the posterior samples. \\

\section{Extending the models} \label{sec:ext}

In the framework of the wrapped and projected normal models, introducing covariate information to explain the angular response
is straightforward.  For the wrapped approach we revise the linear version \eqref{eq:modW} to
$Y(\mathbf{s},t)=\mu_Y(\mathbf{s},t)+{\varepsilon}_Y(\mathbf{s},t)$.
%

The external variables can be introduced by modeling  the mean of the circular process. Linear specification
induces a circular likelihood for the regression coefficients that has infinitely many maxima of comparable size since this
model wraps the line infinitely many times around the circle, \cite[see for example][]{Johnson1978, Fisher1992}.
To address this problem it is customary to  limit the domain of $\mu_Y(\mathbf{s}, t)$ using a link function, i.e.,
$\mu_Y(\mathbf{s}, t)= \mathcal{L}(\mathbf{H}\left(   \mathbf{s},t \right)\boldsymbol{\eta})$, where
$\mathcal{L}(\cdot):\mathbb{R} \rightarrow I$ is the \emph{link function}  and $I$ is some interval of $\mathbb{R}$ of length
equal to the circular variable
period, in our case $2 \pi$. We employ the  \emph{inverse tan link} \citep{Fisher1992}.

If only categorical covariates are available we do not need a link function; we can adopt an ANOVA representation for the
relation between circular response and discrete covariates.  This is computationally more efficient (see Supplementary Online Material, Section S1).
Illustratively, suppose  we have two predictors, with $m_1$ and $m_2$ levels, respectively, say $\mathbf{H}_1=(H_{1,1}, \dots ,H_{1,m_1})$
and $\mathbf{H}_2=(H_{2,1}, \dots , H_{2,m_2})$.  Then, to simplify the condition ensuring $\mu_Y\left(   \mathbf{s},t  \right)  \in I$, we use the following
parametrization:
\begin{equation}
	\mu_Y\left(   \mathbf{s},t  \right)  =
	\sum_{i=1}^{m_1}\sum_{j=1}^{m_2} \mu_{Y,im_2+j} {1}_{\left(   H_1\left(   \mathbf{s},t  \right)=H_{1,i} \right)}{1}_{\left(   H_2\left(   \mathbf{s},t  \right)=H_{2,j} \right)}.
\end{equation}
We can also introduce the covariates into the specifications for the variances, creating $\sigma_Y^2(\mathbf{s},t)$ and
$\phi_Y^2(\mathbf{s},t)$.
Again, we consider ANOVA-type models, e.g., $ \sigma^2_Y\left( \mathbf{s},t \right) =
\sum_{i=1}^{m_1}\sum_{j=1}^{m_2} \sigma^2_{Y,im_2+j} {1}_{\left(   H_1\left( \mathbf{s},t \right)=H_{1,i} \right)}$ ${1}_{\left(
H_2\left(   \mathbf{s},t \right)=H_{2,j} \right)} $ and  $  \phi^2_Y\left( \mathbf{s},t \right) = $
$\sum_{i=1}^{m_1}\sum_{j=1}^{m_2}$ $ \phi^2_{Y,im_2+j} {1}_{\left(   H_1\left(   \mathbf{s},t \right)=H_{1,i} \right)}{1}_{\left(
H_2\left(   \mathbf{s},t \right)=H_{2,j} \right)} $.

We investigate two models, both with an ANOVA parametrization for  $\sigma^2_Y\left(   \mathbf{s},t  \right)$ and $\phi^2_Y\left(
\mathbf{s},t  \right)$ while for the mean, one has  an
ANOVA parametrization  (\emph{WNA}) and the other has a
regression form (\emph{WNR}).  Below, we obtain an ANOVA form if we work with sea state and a regression form if we work
with wave height.
As prior distributions we propose: $N(\cdot,\cdot)$ for $\eta_{Y,i}, i=1,2, \dots$, that is, a customary prior for a regression coefficient; $WrapN(\cdot, \cdot)$ for $\mu_{Y,i}, i = 1,2,\dots$, the circular equivalent of a normal prior over  mean level;
and $IG(\cdot, \cdot)$ for $\sigma_{Y,i}^2$ and $\phi_{Y,i}^2, \, i=1,2, \dots$, that is, the customary prior for a variance. To sample from the predictive distribution we adopt the same procedure used above for the WN model.

To introduce dependence on covariates in the projected normal model, following \cite{Wang2013}, we revise equation \eqref{eq:projz} to
$Z_{\ell}(\mathbf{s}, t) = \mu_{Z_\ell}(\mathbf{s},t)+ \omega_{Z_\ell}(\mathbf{s}, t)+\tilde{\varepsilon}_{Z_\ell}(\mathbf{s},
t), \ell=1,2 $
where the mean of the linear bivariate process is a function of space and/or time and
$\tilde{\varepsilon}_{Z_\ell}(\mathbf{s},t) \stackrel{iid}{\sim} N(0, \phi_Z^2(\mathbf{s},t))$. Then we marginalize over
$\boldsymbol{\omega}_{Z}(\mathbf{s},t)$ to obtain $Z_{\ell}(\mathbf{s}, t) =$ $
\mu_{Z_\ell}(\mathbf{s},t)+{\varepsilon}_{Z_\ell}(\mathbf{s}, t), \ell=1,2$.
We write $\mu_{Z_\ell}(\mathbf{s},t) =$ $ \mathbf{H}(\mathbf{s},t)\boldsymbol{\eta}_{Z_{\ell}}, \ell=1,2$ and $  \phi^2_{Z}\left(
\mathbf{s},t  \right) = \sum_{i=1}^{m_1}\sum_{j=1}^{m_2} \phi^2_{Z,im_2+j}$   $ {1}_{\left(   H_1\left(   \mathbf{s},t
\right)=H_{1,i} \right)}$ ${1}_{\left(   H_2\left(   \mathbf{s},t  \right)=H_{2,j} \right)}$,  where
$\boldsymbol{\eta}_{Z_{\ell}} = (\eta_{Z_{\ell},1},\eta_{Z_{\ell},2}, \dots)^{\prime}$.  Note that, depending on the types of
variables in $\mathbf{H}(\mathbf{s},t)$, continuous or categorical, we can specify a (projected normal) regression (\emph{PNR})
or (projected normal) ANOVA (\emph{PNA}). As noted in \cite{wang2014}, there is complex interaction among the parameters in the
general projected normal, complicating interpretation of the behavior of the resulting projected normal distributions as we vary
them.
With the same rationale used for the priors of the WNA and WNR models, we propose $\eta_{Z_{\ell},i}\sim N(\cdot,\cdot),l=1,2,i=1,2,\dots$ and $\phi_{Z,i} \sim IG(\cdot,\cdot),i=1,2,\dots$.
Here, again, we can sample from the predictive distribution adopting the same procedure as illustrated in Section \ref{sec:predP}.

\subsection{Application to the wave data} \label{sec:real2}
We fitted the new models using the same dataset as in Section \ref{sec:real}. For the ANOVA representation we used, as a categorical
variable, the state of the sea while
for the regression setting we used the significant wave height.
Adopting the same rationale as in  Section \ref{sec:real}, the prior distributions for the regression coefficients  ($\eta_{Y,j,i}$ and $\eta_{Z_{\ell},j,i},j=0,1, i=calm,trans,storm$)
were all $N(0,10)$. For the ANOVA coefficients, $\mu_{Y,i}$ and $\mu_{Z_{\ell},i}$, they were all $WrapN(\pi,10)$. For the  $\sigma_{Y,i}^2$, they were
all $IG(2,2)$ and for the $\phi_{Y,i}$ and $\phi_{Z,i}$ they were all $IG(1,0.25)$. The prior distributions for the other parameters were
the same  as those used in Section \ref{sec:real}.

\begin{figure}[t!]
	\centering
	\subfloat{\includegraphics[trim=0 15 0 50,clip,scale=0.37]{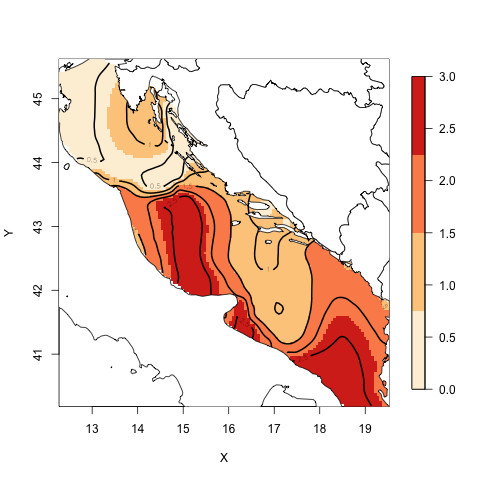}}
	\subfloat{\includegraphics[trim=0 15 0 50,clip,scale=0.37]{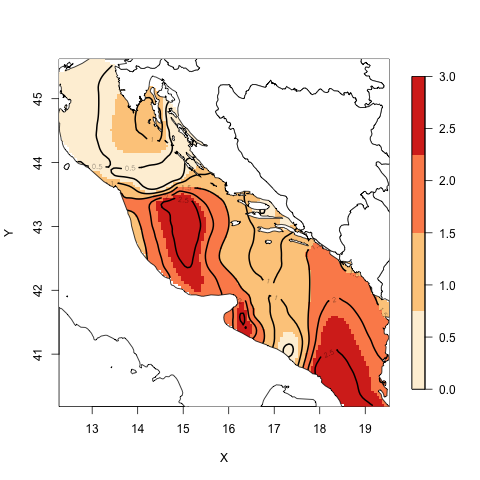}}\\
	\subfloat{\includegraphics[trim=0 15 0 50,clip,scale=0.37]{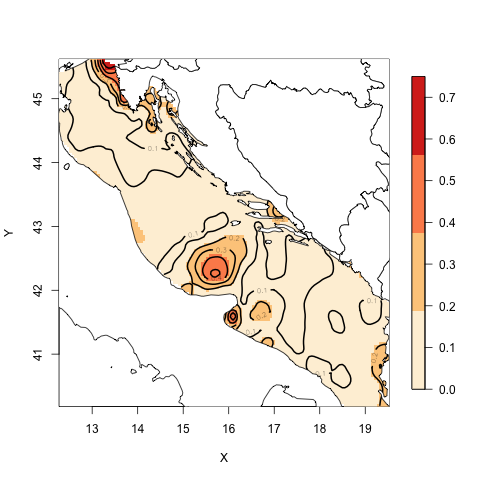}}
	\subfloat{\includegraphics[trim=0 15 0 50,clip,scale=0.37]{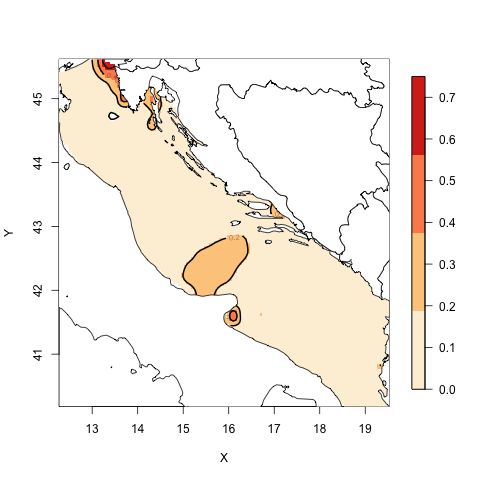}}\\
	\subfloat{\includegraphics[trim=0 15 0 50,clip,scale=0.37]{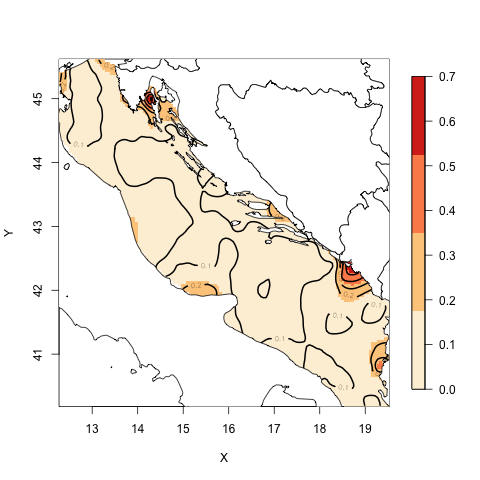}}
	\subfloat{\includegraphics[trim=0 15 0 50,clip,scale=0.37]{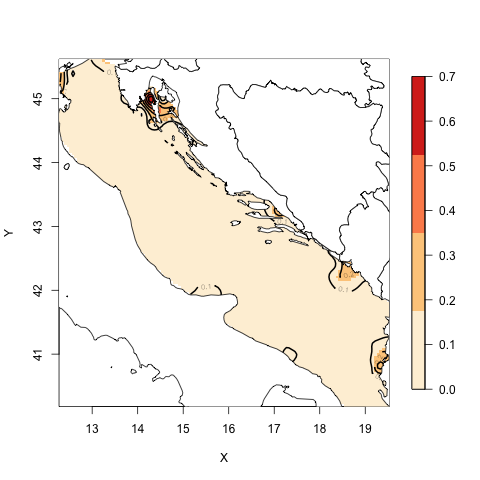}}
	\caption{Real data example: CRPS surfaces for the WN (first column) and PN (second column) models, under calm (first row), transition (second row) and storm (third row) states. Scales differ across states}\label{fig:CRPSsurf}
\end{figure}

\begin{table}[t!]
	\centering
	\caption{Real data example:  CRPS and APE  for WNR, WNA, PNR and PNA models computed on each validation dataset.} \label{tab:CRPSb}
	\begin{tabular}{c|c|c|c|c|c}
		\hline \hline
		& & WNR & WNA& PNR & PNA \\ \hline \hline
		Average &CRPS& 0.668&0.644& 0.507&0.588\\
		& APE & 0.502& 0.431& 0.496&0.450\\\hline
		Calm & CRPS &1.548& 1.409&1.129& 1.342\\
		& APE &1.158& 0.997& 0.985&0.984\\\hline
		Transition & CRPS &0.095& 0.094&0.092&0.093 \\
		& APE &0.033& 0.030&0.046& 0.038\\\hline
		Storm & CRPS &0.057& 0.054&0.118& 0.053\\
		& APE & 0.016& 0.013&0.110& 0.012\\\hline
		One-step prediction & CRPS & 0.971& 1.018&0.689& 0.866\\
		& APE &0.802& 0.685&0.841 & 0.765\\ \hline \hline
	\end{tabular}
\end{table}
From Table \ref{tab:CRPSb} we see that the WNA model is generally preferred to the WNR.  For the projected models, APE and CRPS are almost indistinguishable between PNA and PNR
during transition.
With one-step ahead predictions,
the two criteria return contradicting choices; PNR is preferred with CRPS, PNA with the APE. With the calm sea state, the CRPS
chooses PNR while APE does not yield a clear decision.  With the storm state, both criteria are lower with the PNA model.
%
Overall, our selection would be the PNA model but, more importantly, we
value the informative comparison our approach enables.
In fact, the remarkable improvement of PNA over PNR in storm is likely due to the very high predictability of direction during a
storm period.  In this regard, the PN models are generally preferred to the WN models except in storm where WNR, WNA, and PNA
are essentially equivalent.

To analyze the local behavior of model fitting, in Figure \ref{fig:CRPSsurf} we report CRPS surfaces, evaluated in calm,
transition and storm for the two ``best average APE'' models, the WNA (see Table \ref{tab:CRPSb}) and PN (see Table \ref{tab:CRPS}).
We  see that the local behavior of the models is very similar. The worst predictions are found around the Gargano peninsula during
calm.  This is consistent with the physics of wave movement since, around the peninsula, local winds play a more relevant role,
inducing very high variability in wave directions.
\begin{table}[t!]
	\centering
	\caption{Real data example: mean point estimate (PE) and 95\% credible interval (CI)  for the correlation parameters of the   WNA, WNR, PNA and PNR models} \label{tab:cp3}
	\begin{tabular}{c|c|c|c}
		\hline\hline
		&& WNR & WNA \\ \hline\hline
		$a$ & PE & 0.015&0.008	\\	
		& (CI) &  (0.005,0.035)  &  (0.003,0.020)   \\ \hline
		$c$ & PE & 6.1$\times 10^{-5}$&  4.0$\times 10^{-5}$\\
		& (CI) &   (2.0$\times 10^{-5}$,1.4$\times 10^{-4}$) &  (2.0$\times 10^{-5}$,7.0$\times 10^{-5}$) \\\hline
		$\alpha$ & PE   & 0.620& 0.611 \\
		& (CI) &  (0.445,0.786)& (0.434,0.765)\\\hline
		$\beta$ & PE &  0.396 & 0.539\\
		& (CI) & (0.070,0.830)& (0.181,0.868)\\\hline
		$\gamma$ & (PE) & 0.705 &0.936\\
		& (CI) &  (0.620,0.794)&(0.880,0.976) \\\hline\hline	
		\multicolumn{1}{c}{}&
		\multicolumn{1}{c}{} &
		\multicolumn{1}{c}{} &
		\multicolumn{1}{c}{} \\
		\multicolumn{1}{c}{}&
		\multicolumn{1}{c}{} &
		\multicolumn{1}{c}{} &
		\multicolumn{1}{c}{} \\
		\hline\hline
		&& PNR & PNA \\ \hline\hline
		$a$ & PE & 0.119 &0.108	\\	
		& (CI) &  (0.042,0.267)  &  (0.042,0.225) \\ \hline
		$c$ & PE &  3.0$\times 10^{-3}$ & 1.0$\times 10^{-3}$   \\
		& (CI) &   (1.01$\times 10^{-3}$,1.35$\times 10^{-3}$)    & (4.60$\times 10^{-4}$,3.46$\times 10^{-3}$) \\\hline
		$\alpha$ & PE   & 0.575 & 0.506 \\
		& (CI) & (0.345,0.763 & (0.340,0.706) \\\hline
		$\beta$ & PE & 0.082 &  0.063\\
		& (CI) &  (0.000,0.402) & (0.000,0.300)\\\hline
		$\gamma$ & (PE) & 0.561 & 0.541 \\
		& (CI) &  (0.435,0.677) & (0.441,0.645) \\\hline\hline
	\end{tabular}
\end{table}
 The same behavior is shown with the other models. In terms of parameter
estimation the WNA and PN models suggest a non-separable model (Tables \ref{tab:cp3} and \ref{tab:cp1}) with very strong spatial
(${c}$) and temporal (${a}$) dependence. WNA  suggests that a different nugget is necessary
for each sea state. In fact analyzing the credible intervals of these parameters we observe that, for each sea state, nuggets are significantly different among them as their credible intervals do not overlap  (Table  \ref{tab:WNrd2}). For the projected normal models (Table \ref{tab:PNrd2}), all
nugget credible intervals are substantially overlapping,  suggesting that one nugget should be enough to model all sea states.
\begin{table}[t]
	\centering
	\caption{  Real data example: mean point estimate (PE) and 95\% credible interval (CI) of the parameters of the WNA and WNR models.
	}\label{tab:WNrd2}
	{
		\begin{tabular}{c|ccccc}
			\multicolumn{1}{c}{}&
			\multicolumn{1}{c}{} &
			\multicolumn{1}{c}{WNA} &
			\multicolumn{1}{c}{} \\
			\hline\hline
			&	${\mu}_{Y,calm}$  &  ${\sigma}_{Y,calm}^2$ & ${\phi}_{Y,calm}^2$  & \\
			\hline
			PE & 0.095 & 1.524 & 0.051  \\
			(CI)& (5.232,1.328)  &  (0.959,2.387)&(0.039,0.068)    \\ \hline\hline
			&	${\mu}_{Y,tran}$  &  ${\sigma}_{Y,tran}^2$ & ${\phi}_{Y,tran}^2$  \\
			\hline
			PE &5.998   & 0.541 &  0.018  \\
			(CI) &(5.278,0.490)  &  (0.332,0.876)&(0.013,0.026)    \\ \hline \hline
			&	${\mu}_{Y,storm}$  &  ${\sigma}_{Y,storm}^2$ & ${\phi}_{Y,storm}^2$  \\
			\hline
			PE &5.860  & 0.385 &  0.009  \\
			(CI)& (5.254,0.281)  &  (0.246,0.582)&(0.007,0.012)    \\   \hline\hline
			\multicolumn{1}{c}{}&
			\multicolumn{1}{c}{} &
			\multicolumn{1}{c}{} &
			\multicolumn{1}{c}{} \\
			
			\multicolumn{1}{c}{}&
			\multicolumn{1}{c}{} &
			\multicolumn{1}{c}{WNR} &
			\multicolumn{1}{c}{} \\
			\hline\hline
			&	${\eta}_{Y,0,calm}$ & ${\eta}_{Y,1,{calm}}$    &  ${\sigma}_{Y,calm}^2$ & ${\phi}_{Y,calm}^2$   \\
			\hline
			PE &0.997& 4.918&5.000&0.041   \\
			(CI)&	(0.360,1.901)  &(2.433,7.619)  &(2.313,9.494)  &(0.027,0.058)       \\ \hline\hline
			&	${\eta}_{Y,0,tran}$  & ${\eta}_{Y,1,{tran}}$&  ${\sigma}_{Y,tran}^2$ & ${\phi}_{Y,tran}^2$   \\
			\hline
			PE &3.166 &2.526&1.825  &0.018  \\
			(CI)&	(0.763,5.894)  &(0.174,6.844)  &(1.013,3.046)  &(0.012,0.025)      \\ \hline\hline
			&${\eta}_{Y,0,storm}$ & ${\eta}_{Y,1,{storm}}$ &  ${\sigma}_{Y,storm}^2$ & ${\phi}_{Y,storm}^2$  \\
			\hline
			PE &3.470 &1.933&1.322&0.010  \\
			(CI)&(0.666,6.445)  &(0.064,5.870)  &(0.660,2.167)  &(0.007,0.013)      \\   \hline\hline
		\end{tabular}
	}
\end{table}

\begin{table}[t]
	\centering
	\caption{Real data example: mean point estimate (PE) and 95\% credible interval (CI)  for the parameters of the  PNA and PNR models.
	}\label{tab:PNrd2}
	{
		\begin{tabular}{c|ccccc}
			\multicolumn{1}{c}{}&
			\multicolumn{1}{c}{} &
			\multicolumn{1}{c}{PNA} &
			\multicolumn{1}{c}{} \\
			\hline\hline
			&${\mu}_{Z_1,calm}$ &  ${\mu}_{Z_2,calm}$      & ${\phi}_{Z,calm}^2$   \\
			\hline
			PE & 0.841 &-0.404 & 0.027  \\
			(CI)&(-1.112,2.706)& (-2.408,1.427)& (0.014,0.051)    \\ \hline\hline
			&	${\mu}_{Z_1,tran}$ &  ${\mu}_{Z_2,tran}$ &   ${\phi}_{Z,tran}^2$   \\
			\hline
			PE &0.697 &-0.724 &0.047    \\
			(CI)&(-1.281,2.640)& (-2.600,1.173)& (0.018,0.099)    \\ \hline\hline
			&	${\mu}_{Z_1,storm}$ &  ${\mu}_{Z_2,storm}$ &  ${\phi}_{Z,storm}^2$  \\
			\hline
			PE &0.615 &-0.615 &0.037   \\
			(CI)&(-1.376,2.615)& (-2.543,1.289)& (0.016,0.076)      \\   \hline\hline
			&${\sigma}_{Z,1}^2$ & ${\rho}_{Z}$     \\
			\hline
			(PE) &	2.072	&-0.161\\
			(CI)&(1.425,2.938)& (-0.320,0.003)\\	\hline \hline
			\multicolumn{1}{c}{}&
			\multicolumn{1}{c}{} &
			\multicolumn{1}{c}{} &
			\multicolumn{1}{c}{} \\
			\multicolumn{1}{c}{}&
			\multicolumn{1}{c}{} &
			\multicolumn{1}{c}{PNR} &
			\multicolumn{1}{c}{} \\ 			
			%
			%
			\hline\hline
			&${\eta}_{Z_1,0,calm}$ &   ${\eta}_{Z_1,1,{calm}}$ &  ${\eta}_{Z_2,0,calm}$      & ${\eta}_{Z_2,1,{calm}}$ & ${\phi}_{Z,calm}^2$    \\
			\hline
			PE & 0.997&  0.875& -0.925  &0.840 & 0.110  \\
			(CI)  &	(-0.989,3.026)&(-1.160,2.927)&(-2.878,1.091)&(-1.162,2.798)&(0.033,0.250)    \\ \hline\hline
			&	${\eta}_{Z_1,0,tran}$ &   ${\eta}_{Z_1,1,{tran}}$ &  ${\eta}_{Z_2,0,tran}$      & ${\eta}_{Z_2,1,{tran}}$ &   ${\phi}_{Z,tran}^2$   \\
			\hline
			PE & 	0.916 & 0.976  &-1.117  & -0.554 & 0.127   \\
			(CI)  &	(-1.195,3.015)&(-1.258,3.117)&(-3.322,0.893)&(-2.601,1.649)&(0.037,0.322)   \\ \hline\hline
			&	${\eta}_{Z_1,0,storm}$ &   ${\eta}_{Z_1,1,{storm}}$ &  ${\eta}_{Z_2,0,storm}$      & ${\eta}_{Z_2,1,{storm}}$ &  ${\phi}_{Z,storm}^2$  \\
			\hline
			PE & 	0.768  & 1.088 & -0.974 & -1.190 & 0.091   \\
			(CI)  &	(-1.424,2.899)&(-1.083,3.235)&(-3.146,1.177)&(-3.281,0.955)&(0.031,0.201)   \\   \hline\hline
			&	${\sigma}_{Z,1}^2$ &  ${\rho}_{Z}$ \\
			\hline
			PE & 	2.293   &-0.191 \\
			(CI)  &	(1.602,3.212) &(-0.358,-0.013) \\   \hline\hline
		\end{tabular}
	}
\end{table}



\section{Conclusions}\label{sec:4} \label{sec:conc}
We have presented a range of models for spatio-temporal circular data based on  the wrapped and projected normal distributions,
incorporating space-time dependence, allowing explanatory variables, introducing a nugget, implementing kriging and forecasting.
The models based on the projected normal are more flexible since they allow bimodal and asymmetric distributions while the
wrapped normal is unimodal and symmetric.  On the other hand, the wrapped normal models are easy to interpret and are computationally better
behaved and more efficient.
Predictions obtained under the two models are very close  and almost indistinguishable when data are roughly unimodal and
symmetric (see Supplementary Online Material,  Section  S2). Then, if fast computation is sought, WN models become attractive.

The projected normal process can  be straightforwardly extended to general directional fields on the  sphere since the projected
normal distribution is well defined in this case, see \cite{Merdia1999}. The wrapped Gaussian  process is not easily extended to a
sphere.  In fact, we are unaware of any approaches to wrap multivariate linear data onto spheres. Conceptually, such wrapping
would \emph{not} appear to be well defined.

Future work will find us enriching wrapped  modeling to allow asymmetry through the use of skewed distributions. Skewness
is easy to introduce by wrapping  skew normal distributions. In a completely different direction,
we are also extending the modeling to explore spatio-temporal data consisting of geo-coded locations with periodic (in time) behaviour that can be represented as a circular variable. There, we work with trivariate GP's in
space and time, incorporating temporal projection. 

\section*{Acknowledgement}
The authors thank  INFN Bari CED for allowing  the use of their high performance grid computing infrastructure Bc2S. The
authors thank ISPRA for the use of data output from the wave
model of its SIMM hydro-meteo-marine forecasting system.

\bibliographystyle{WPspbasic}      
\bibliography{WPbib}   

\begin{thebibliography}{28}
\providecommand{\natexlab}[1]{#1}
\providecommand{\url}[1]{{#1}}
\providecommand{\urlprefix}{URL }
\expandafter\ifx\csname urlstyle\endcsname\relax
  \providecommand{\doi}[1]{DOI~\discretionary{}{}{}#1}\else
  \providecommand{\doi}{DOI~\discretionary{}{}{}\begingroup
  \urlstyle{rm}\Url}\fi
\providecommand{\eprint}[2][]{\url{#2}}

\bibitem[{Banerjee et~al(2014)Banerjee, Gelfand, and Carlin}]{Banerjee2003}
Banerjee S, Gelfand AE, Carlin BP (2014) Hierarchical modeling and analysis for
  spatial data, 2nd edn. Chapman and Hall/CRC, New York

\bibitem[{Breckling(1989)}]{breckling1989}
Breckling J (1989) The analysis of directional time series: applications to
  wind speed and directions. Lecture notes in statistics, Springer-Verlag,
  Berlin

\bibitem[{Bulla et~al(2012)Bulla, Lagona, Maruotti, and Picone}]{Bulla2012}
Bulla J, Lagona F, Maruotti A, Picone M (2012) A multivariate hidden {M}arkov
  model for the identification of sea regimes from incomplete skewed and
  circular time series. J Agr Biol Environ Stat 17:544--567

\bibitem[{Coles(1998)}]{coles98}
Coles S (1998) Inference for circular distributions and processes. Stat Comput
  8:105--113

\bibitem[{Damien and Walker(1999)}]{damien-walker}
Damien P, Walker S (1999) A full {B}ayesian analysis of circular data using the
  von {M}ises distribution. Can J Stat 27:291--298

\bibitem[{Fisher(1996)}]{fisher1996}
Fisher NI (1996) Statistical analysis of circular data. Cambridge University
  Press, Cambridge

\bibitem[{Fisher and Lee(1992)}]{Fisher1992}
Fisher NI, Lee AJ (1992) Regression models for an angular response. Biometrics
  48:665--677

\bibitem[{Gneiting(2002)}]{Gneiting2002}
Gneiting T (2002) Nonseparable, stationary covariance functions for space--time
  data. J Am Stat Assoc 97:590--600

\bibitem[{Grimit et~al(2006)Grimit, Gneiting, Berrocal, and
  Johnson}]{grimit2006}
Grimit EP, Gneiting T, Berrocal VJ, Johnson NA (2006) The continuous ranked
  probability score for circular variables and its application to mesoscale
  forecast ensemble verification. Q J R Meteorol Soc 132:2925--2942

\bibitem[{Guttorp and Lockhart(1988)}]{guttorp1998}
Guttorp P, Lockhart RA (1988) Finding the location of a signal: A {B}ayesian
  analysis. J Am Stat Assoc 83:322--330

\bibitem[{Harrison and Kanji(1988)}]{harrison}
Harrison D, Kanji GK (1988) The development of analysis of variance for
  circular data. J Appl Stat 15:197--224

\bibitem[{Holtzman et~al(2006)Holtzman, Munk, Suster, and
  Zucchini}]{Holzmann2006}
Holtzman H, Munk A, Suster M, Zucchini W (2006) Hidden {M}arkov models for
  circular and linear-circular time series. Environ Ecol Stat 13:325--347

\bibitem[{Jammalamadaka and SenGupta(2001)}]{Jammalamadaka2001}
Jammalamadaka SR, SenGupta A (2001) Topics in circular statistics. World
  Scientific, Singapore

\bibitem[{Johnson and Wehrly(1978)}]{Johnson1978}
Johnson RA, Wehrly TE (1978) Some angular-linear distributions and related
  regression models. J Am Stat Assoc 73:602--606

\bibitem[{Jona~Lasinio et~al(2012)Jona~Lasinio, Gelfand, and
  Jona~Lasinio}]{Jona2013}
Jona~Lasinio G, Gelfand AE, Jona~Lasinio M (2012) Spatial analysis of wave
  direction data using wrapped {G}aussian processes. Ann Appl Stat 6:1478--1498

\bibitem[{Kato(2010)}]{kato2010}
Kato S (2010) A {M}arkov process for circular data. J Roy Statist Soc Ser B
  72:655--672

\bibitem[{Kato and Shimizu(2008)}]{Katoa2008}
Kato S, Shimizu K (2008) Dependent models for observations which include
  angular ones. J Stat Plan Infer 138:3538--3549

\bibitem[{Lagona and Picone(2011)}]{lagona2011}
Lagona F, Picone M (2011) A latent-class model for clustering incomplete linear
  and circular data in marine studies. J Data Sci 9:585--605

\bibitem[{Lee(2010)}]{lee2010}
Lee A (2010) Circular data. Wiley Interdiscip Rev Comput Stat 2:477--486

\bibitem[{Mardia(1972)}]{mardia72}
Mardia KV (1972) Statistics of directional data. Academic Press, London

\bibitem[{Mardia and Jupp(1999)}]{Merdia1999}
Mardia KV, Jupp PE (1999) Directional statistics. John Wiley and Sons,
  Chichester

\bibitem[{Mastrantonio et~al(2015)Mastrantonio, Maruotti, and
  Jona~Lasinio}]{mastrantonio2015}
Mastrantonio G, Maruotti A, Jona~Lasinio G (2015) Bayesian hidden {M}arkov
  modelling using circular-linear general projected normal distribution.
  Environmetrics 26:145--158

\bibitem[{Ravindran and Ghosh(2011)}]{ravindran}
Ravindran P, Ghosh SK (2011) Bayesian analysis of circular data using wrapped
  distributions. Stat Sinica 5:547--561

\bibitem[{Speranza et~al(2004)Speranza, Accadia, Casaioli, Mariani, Monacelli,
  Inghilesi, Tartaglione, Ruti, Carillo, Bargagli, Pisacane, Valentinotti, and
  Lavagnini}]{speranza2004}
Speranza A, Accadia C, Casaioli M, Mariani S, Monacelli G, Inghilesi R,
  Tartaglione N, Ruti PM, Carillo A, Bargagli A, Pisacane G, Valentinotti F,
  Lavagnini A (2004) Poseidon: An integrated system for analysis and forecast
  of hydrological, meteorological and surface marine fields in the
  {M}editerranean area. Nuovo Cimento 27:329--345

\bibitem[{Speranza et~al(2007)Speranza, Accadia, Mariani, Casaioli,
  Tartaglione, Monacelli, Ruti, and Lavagnini}]{MET:MET34}
Speranza A, Accadia C, Mariani S, Casaioli M, Tartaglione N, Monacelli G, Ruti
  PM, Lavagnini A (2007) Simm: an integrated forecasting system for the
  {M}editerranean area. Meteorol Appl 14:337--350

\bibitem[{Wang and Gelfand(2013)}]{Wang2013}
Wang F, Gelfand AE (2013) Directional data analysis under the general projected
  normal distribution. Stat Methodol 10:113--127

\bibitem[{Wang and Gelfand(2014)}]{wang2014}
Wang F, Gelfand AE (2014) Modeling space and space-time directional data using
  projected {G}aussian processes. J Am Stat Assoc 109:1565--1580

\bibitem[{Wang et~al(2015)Wang, Gelfand, and Jona~Lasinio}]{wang:2014}
Wang F, Gelfand AE, Jona~Lasinio G (2015) Joint spatio-temporal analysis of a
  linear and a directional variable: space-time modeling of wave heights and
  wave directions in the {A}driatic {S}ea. Stat Sinica 25:25--39

\end{thebibliography}


\end{document}